\documentclass{aa}
\usepackage{times,txfonts,latexsym,graphics,graphicx,lscape,longtable}
\usepackage[dvips]{color}

\usepackage{natbib}
\bibpunct{(}{)}{;}{a}{}{,}

\newcommand{\teff}{$T_{\rm eff}$}
\newcommand{\logg}{$\log g$}
\newcommand{\vmic}{$\xi_{\rm t}$}

\begin{document}

\title{Abundances and possible diffusion of elements in M67 stars}
\author{Anna \"Onehag \inst{1} 
\and    Bengt Gustafsson \inst{1,2} 
\and    Andreas Korn \inst{1}
}

\authorrunning{\"Onehag et al.}

\institute{Department of Physics and Astronomy, 
           Uppsala Astronomical Observatory,
           Box 515, SE-751\,20 Uppsala, Sweden
\and Nordita, Roslagstullsbacken 23, SE-106\,91 Stockholm, Sweden
}

\date{Received 13 September 2013 / Accepted 25 November 2013}
\offprints{Anna \"Onehag,,
\email{Anna.Onehag@physics.uu.se}}

\authorrunning{\"Onehag et al.}
\titlerunning{Abundances and diffusion in M67 stars}

\abstract
{
The rich open cluster M67 is known to have a chemical
composition close to solar and an age of about 3.5--4.8\,Gyr. It
offers an important opportunity to check and develop our understanding
of the physics and the evolution of solar-type stars.
}
{
We present a spectroscopic study at high resolution, $R \approx$ 
50,000, of 14 stars located on the main sequence, at the turn-off point,
and on the early subgiant branch in the cluster in order to investigate
its detailed chemical composition, for comparison with the Sun and
solar twins in the solar neighbourhood, and to explore selective atomic diffusion
of chemical elements as predicted by stellar-structure theory.
}
{We have obtained VLT/FLAMES-UVES spectra and analysed these strictly
differentially in order to explore chemical-abundance similarities and differences
between the M67 stars and the Sun and among the M67 stars themselves.
}
{Individual abundances of 19 different chemical elements are obtained for the stars.
They are found to agree very well with solar abundances, with abundance
ratios closer to solar than those of most solar twins in the solar neighbourhood.
An exception is Li, which shows considerable scatter among the cluster stars.
There is a tendency for the cluster-star abundances to be more depleted than the abundances
in the field stars in correlation with the
condensation temperature of the elements, a tendency also found earlier 
for the Sun. Moreover, the heavy-element abundances
are found to be reduced in the hotter stars and dwarfs by typically $\leq$0.05\,dex, as compared
to the abundances of the subgiants.  
 }
{The results support the hypothesis that the gas of the proto-cluster was depleted by
formation and cleansing of dust before the stars formed. They also add support to the proposal
that the Sun was formed in a dense stellar environment. Moreover, the observed minor reductions 
of heavy elements, relative to our standard star M67-1194 and the subgiants, 
in the atmospheres of dwarfs and turn-off point stars seem to suggest that diffusion processes 
are at work in these stars, although the evidence is not compelling. Based on theoretical 
models, the diffusion-corrected initial metallicity of M67 is estimated to be [Fe/H]=+0.06.
 }


\maketitle

\section{Introduction}
\label{sec:intro}

The old and rich open cluster M67 offers interesting possibilities for
studying the evolution of solar-like stars. 
The cluster has a solar-similar chemical composition with
[Fe/H] in the range $-$0.04 to +0.03 \citep{Hobbs&Thorburn:91,Tautvaisiene:00,
Yong&al:05,Randich&al:06,Pace&al:08,Pasquini&al:08}.
Its age is also comparable to the Sun: 3.5--4.8\,Gyr \citep{Yadav&al:08}.
M67 is relatively nearby \citep[$\sim$\,800-900\,pc,][]{Majaess&al:11, Sarajedini&al:09, Yakut&al:09} 
and is affected only a little by interstellar extinction,
which allows for detailed spectroscopic studies of even its main-sequence stars. 

M67 thus seems to offer good possibilities of finding solar-twin  
candidates for further exploration. \citet{Pasquini&al:08} 
\citep[followed by a paper of][]{Biazzo&al:09} have listed ten promising 
twin candidates in the cluster. We have previously analysed one of these spectroscopically
\citep[M67--1194,][]{Onehag&al:11} and found it to have a chemical composition
very close to solar. This is noteworthy since \citet{Melendez&al:09} and
\citet{Ramirez&al:09} have systematically analysed solar-twin candidates
in the solar neighbourhood  and found that, although these stars in general have  
fundamental parameters very close to solar, 
differential high-accuracy abundance analyses prove that almost all of them have chemical
compositions systematically {\it deviating} from that of the Sun: the
twins are, in relative terms, slightly richer in refractory elements (that condense at high temperatures)
than volatiles.  Unlike these nearby solar twins, the composition of M67-1194 is more 
solar-like, which lends some support to the idea that the Sun was once formed in a similar 
cluster environment. It is of strong interest to investigate whether analyses of other
stars in M67 verify this remarkable solar likeness.

A cluster like M67 also offers the possibility of exploring the changes in surface composition
of solar-type stars as a result of selective diffusion of elements in the stars. Such effects 
are expected to deplete heavy elements from the surface, but to different degrees for different 
elements depending on the element-specific radiative levitation. Effects of this nature  
have been found by \citet{Korn&al:07}, \citet{Lind&al:08}, and \citet{Nordlander&al:12} for the
metal-poor globular cluster NGC~6397. These authors compared the abundances of
turn-off point stars, subgiants, and red giants, where the cooler stars are mixed by the deeper 
convection zone and thus are less affected, and found satisfactory agreement with theoretical predictions 
of atomic diffusion moderated by surface convection and (parametrised) additional mixing of unknown origin.  
Similar effects, albeit of smaller overall amplitude, have recently been traced in NGC~6752 at somewhat 
higher metallicities \citep{Gruyters&al:13}. In M~4, at roughly one tenth solar metallicity, 
no abundance trend in iron was found between the turn-off point and the red giants \citep{Mucciarelli&al:11}, 
possibly indicating that the more massive outer convection zones of more metal-rich stars suppress atomic 
diffusion.
 
It is now important to find out to what extent such diffusion effects are also visible in dwarf stars
of solar metallicity. This is particularly important since solar-like stars are used as the main tracers 
of the chemical evolution in the Galaxy, and any physical effects that change their surface elemental 
composition during their evolution must then be taken into consideration in detailed work on Galactic 
evolution. 
 
Here we present the analysis of 14 stars in M67, six main-sequence (MS) stars (including M67-1194, which 
is used as a standard star in this study), three stars at the turn-off point (TO), and five
on the early subgiant branch (SG). The analysis is based on high-resolution observations with relatively high 
signal-to-noise (S/N) ratio. In Sect.~\ref{sec:obs}, we present the observations and the data 
reduction. Sect.~\ref{sec:analys} describes the analysis method and the estimation of fundamental 
parameters  (\teff, \logg, [Fe/H], and \vmic). In Sect.~\ref{sec:comp} we present the results of a 
detailed analysis of a number of chemical elements and discuss the results.
In Sect.~\ref{sec:conclusions} we summarise our main results.

\section{Observations}
\label{sec:obs}
The observations of M67-1194 were carried out with the multi-object spectrograph FLAMES/UVES at ESO-VLT 
UT2 in Service Mode in the spring of 2009 during a period of three months (18th of January -- 3rd of April,
project 082.D-0726(A), P.I. Gustafsson). The observations were arranged such that approximately a similar
number of photons were collected for each star, resulting in longer total exposure times for the fainter stars.
In each observing block of the observations,  one fibre of the spectrograph system was positioned on
M67-1194 in order to collect as many observations as possible of this faintest star in the programme.

We obtained altogether 23 individual observations in 13 observing nights (18\, h of net exposure time). 
The chosen spectrograph setting (RED580) yields a resolution  of 
$R = \lambda/\Delta\lambda=$\,47,000 (1'' fibre) and a wavelength coverage of 4800--6700\,\AA. A typical
signal-to-noise ratio (S/N) per frame of 50 per unbinned pixel was achieved. Baryocentric radial velocities 
were determined from the individual spectra indicating a radial velocity of $33 \pm 1$\,kms$^{-1}$.
This is in excellent agreement with the mean radial velocity of the cluster as 
determined by \citet{Yadav&al:08}, giving $33.67 \pm 0.09$\,kms$^{-1}$. The frames were subsequently 
co-added for highest possible S/N ratio.

\subsection{Data reduction}
While ESO provides pipeline-reduced data for an initial assessment of the
observed spectra, the resulting data are not intended for a full scientific analysis. In
fact, we found that the extraction of our spectra by the ESO pipeline was
problematic, mainly because the instrumental setup of FLAMES/UVES combined with
our choice of targets is challenging. Having eight fibres tightly packed in a
slit-like fashion, FLAMES/UVES can obtain echelle spectra of up to 8 
targets in parallel. However, the spacing between the individual fibres is 
rather small, which leads to non-negligible light contamination between 
adjacent fibres when the apparent magnitudes of the targets differ significantly. 
This was the case, as M67-1194 is considerably fainter than any of the other stars
placed on one of the adjacent fibres. The ESO 
pipeline had severe problems extracting such a spectrum, which is apparent 
from a resulting strong and periodic pattern of low-sensitivity spikes.  These difficulties
were overcome using a special reduction procedure, based on the echelle-reduction package
REDUCE \citep{Piskunov&Valenti:02}, and further described in \citet{Onehag&al:11}. 
In this work we benefited from important contributions by Dr.\ Eric Stempels. 
The periodic pattern of spikes was removed successfully, and the agreement with the 
pipeline-reduced spectra outside the spikes is excellent. The S/N in 
the co-added spectra of the program star is typically 150 per rebinned pixel,
where a rebinning by two has been applied retaining the full resolution of FLAMES-UVES.

\section{Analysis}
\label{sec:analys}

\subsection{Fundamental stellar parameters}
\label{sec:stellarparam}

{\it Effective temperatures:} In establishing the effective temperatures of the programme stars 
we have used the colours ($V-K_{\rm s}$) and ($V-I_{\rm c}$), as well as the wings of the 
H$\alpha$ lines, as primary indicators. We preferred the use 
of colours and H$\alpha$-line profiles to the use of metal-lines of different excitation and ionisation. The
equilibria of atoms and ions are known to be affected by departures from LTE, 
and since we aimed at exploring small differences in abundances between
stars with different surface gravities (i.e. atmospheric pressures) and 
effective temperatures, we were anxious to avoid systematic errors due to the 
problems of modelling such non-LTE effects and their pressure and temperature dependencies 
(in particular the effect of inelastic collisions with neutral hydrogen
as well as the exact temperature sensitivity of the photo-ionising ultraviolet flux) 
from first principles. However, in checking our temperature scale we investigated the 
temperatures resulting also from the excitation of Fe\,I and the ionisation of Fe and Ti, 
see below. 

The ($V-K_{\rm s}$) and ($V-I_{\rm c}$) colours were adopted from the 2MASS Catalogue \citep{Cutri&al:03}
and \citet{Yadav&al:08}, respectively, with K$_s$ measured in the 2MASS and I$_c$ measured in the 
Johnson-Cousins system. A reddening $E(B-V)$ of 0.041\,$\pm$\,0.004 \citep{Taylor:07}
was adopted and the \teff\ calibrations of \citet{Casagrande&al:10} applied. In these
calibrations neither differences in \logg\ among the stars at or near the main sequence, nor the
effects of varying microturbulence parameters were considered. We verified that these effects only 
correspond to $\Delta T_{\rm eff} \sim $30$\, {\rm K/dex}\, \times\, \Delta \log g$ 
and  $\Delta T_{\rm eff} \sim -5 {\rm K/km\,s^{-1}}\, \times\, \Delta \xi_{\rm t}$ by
model-atmosphere calculations performed with the MARCS program \citep{Gustafsson&al:08}. 
In practice, these corrections can thus be neglected. 

\begin{figure}[htbp]
  \centering
   \resizebox{\hsize}{!}
    {\includegraphics[angle=90]{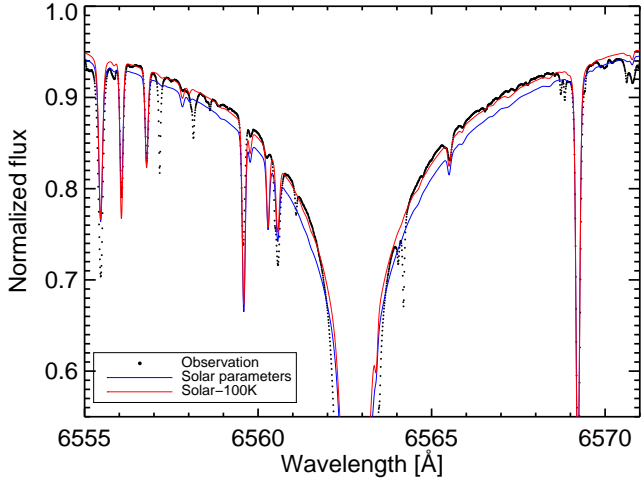}}
  \caption{Part of the observed solar spectrum (from the Kitt Peak Atlas, black dots) and the SME 
           synthetic spectrum from a MARCS model with solar parameters (blue), as well as a MARCS model 
           with \teff\,=\,5670\,K, else solar parameters (red). Obviously, the MARCS solar model produces a too
           wide H$\alpha$ line.}
  \label{F1}
\end{figure}

\begin{figure}[htbp]
  \centering
   \resizebox{\hsize}{!}
    {\includegraphics[angle=90]{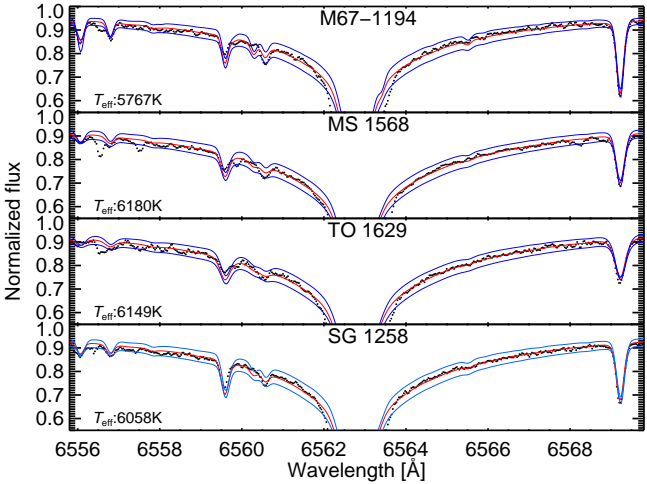}}
  \caption{Fits of the H$\alpha$ wing profiles of four representative stars. 
           The red lines show spectra of best fit models. Blue lines show the effects of 
           temperature shifts by $\pm$\,100\,K. Effective temperatures given in the
           panels are corrected upwards by 100\,K (see text).}
  \label{F2}
\end{figure}

In fitting the wings of the H$\alpha$ lines we used the synthetic spectrum programme SME
\citep{Valenti&Piskunov:96,Valenti&Fischer:05} and
MARCS models \citep{Gustafsson&al:08}. SME uses the H$\alpha$ line-broadening data of \citet{Barklem&al:00}.
We started by comparing the wings calculated using a MARCS \teff/\logg/[Fe/H] = 5780K/4.44/0.0
solar model with the Kitt Peak Atlas \citep{Kurucz:84} of the observed solar flux. The result
is shown in Fig.~\ref{F1}. Obviously, the calculated profile is too wide, and a good
solar fit would require a model temperature around 5670\,K. Assuming that models for slightly
hotter stars would suffer from a similar mismatch to observations, we have schematically
increased the temperatures derived from the H$\alpha$ fits by 100\,K. This result agrees very well
with that of \citet{Cayrel&al:11} who using ATLAS model atmospheres found an effective temperature
from H$\alpha$ 99\,K too cool for the Sun and a systematic shift of this magnitude for solar-type stars
with measured angular diameters. (We note, however, that \citet{Cornejo&al:12}, using a similar method,
find a solar temperature from H$\alpha$ fits that is only 30\,K too low.
We also note that an independent attempt to fit the H$\alpha$ profiles with the less detailed MAFAGS-ODF 
model atmospheres \citep{Grupp:04} and the outdated broadening data of 
\citet{Ali&Griem:65} leads to discrepancies with the observed solar wing profile that are much smaller.)
Examples of our H$\alpha$ fits are shown in Fig.~\ref{F2}. 
We have also made similar analyses of the H$\beta$ line which supports a temperature correction
of 100\,K. The \teff\ values resulting from fits of H$\beta$ wings 
show an increased scatter when compared with results from colours and other temperature indicators (see below), 
which we ascribe to the less favourable placement of H$\beta$ towards the edge of two echelle orders and the 
higher density of blending metal lines in this part of the spectrum.

\begin{figure}[htbp]
  \centering
   \resizebox{\hsize}{!}
    {\includegraphics[angle=90]{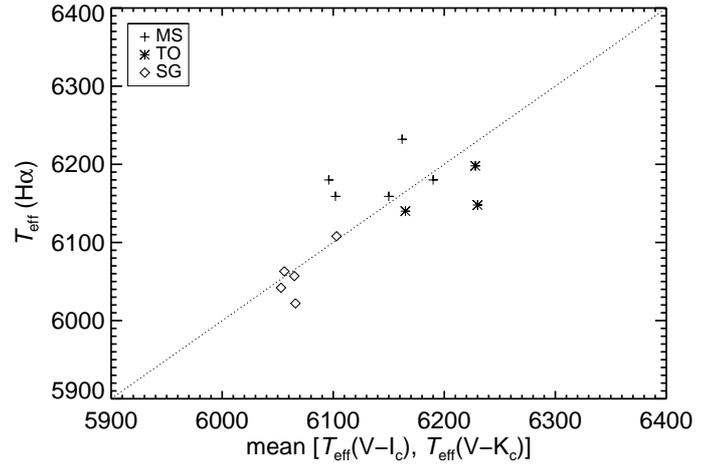}} 
  \caption{The temperature estimates from the H$\alpha$ profiles, corrected by +100\,K, plotted vs the 
           mean of the temperatures derived from the ($V-I_{\rm C}$) and ($V-K_{\rm S}$) colours. 
           Programme stars of different types are represented by different symbols. The dotted 
           line is a 1-1 line.}
  \label{F3}
\end{figure}

\begin{figure}[htbp]
  \centering
   \resizebox{\hsize}{!}
    {\includegraphics[angle=90]{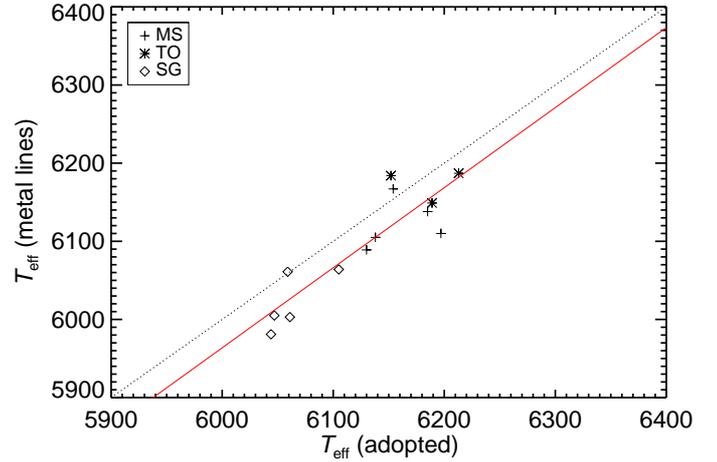}}
  \caption{The mean of the temperature estimates from ionisation equilibria (Fe and Ti, respectively),
           excitation equilibria and relative line depths of metal lines, vs the adopted effective temperatures. 
           The 1-1 line (dots) and a regression line (red) are drawn. Symbols for different stellar groups are 
           indicated.}           
  \label{F4}
\end{figure}

The resulting mean difference  between the two photometric determinations \teff ($V-K_{\rm s}$) and 
\teff ($V-I_{\rm c}$) is $ 14\,K\pm 48\,K$ (s.d). In Fig.~\ref{F3} we have plotted the temperatures 
determined from H$\alpha$ vs the means of the two photometric temperatures with the correction mentioned above.
It is seen that these two temperature scales agree well. The mean difference is $-$2\,K and the standard deviation 
is 46\,K. There may be a tendency for the different stellar groups to behave somewhat differently, the 
MS stars being located higher in Fig.~{\ref{F3}} than the SG and TO stars. Such systematic differences 
may reflect systematic errors in the relative temperatures and the effects of these on the abundances 
derived will be discussed below. {\it We have adopted a mean of these two scales, the one based on colours 
and that based on H$\alpha$ wings as temperature estimates for the programme stars.}

In checking these temperatures, we have compared with ionisation temperatures, obtained
from Fe I and Fe II lines, as well as Ti I and Ti II lines. 
The neutrals and the corresponding ions should give identical elemental abundances in the analyses. A good
agreement is found, although with a zero-point shift of  $61{\rm\,K} \pm{\rm 49\,K}$ (s.d.).  
One should note that these ionisation temperatures are essentially based on very few unsaturated lines from
Fe\,II and Ti\,II, respectively. We have also compared with
temperatures obtained by requiring identical abundances resulting from Fe I 
lines at different excitation. Here, we find a mean difference between the adopted 
temperatures and those obtained from the excitation equilibria in LTE of $ 25\,{\rm K} \pm 86\, $K (s.d.). 
We have finally also checked the temperature scale
by using the method of line-depth ratios of \citet{Gray:94} for lines from the same element with 
different excitation energies. Specifically, we selected the line pair Fe I $\lambda 5679$ and FeI 
$\lambda 6498$, and fitted the macro-turbulence parameter individually for each star, using a number
of reasonably clean lines in the spectral interval to mimic the combination of effects of macroscopic 
motions in the atmospheres, rotation and instrumental profiles. We derived in this way effective temperatures 
that depart at a mean from the adopted ones by $-15\,{\rm K} \pm50\,$K (s.d.). 

In Fig.~\ref{F4} we have plotted the mean of the spectroscopic temperature estimates, from ionisation equilibria
(Fe and Ti respectively), excitation equilibria (of Fe\,I), and relative line depths (Fe\,I) 
relative to the adopted temperatures. The two scales are compatible at the 33\,$\pm$\,32\,K level.

Altogether, we estimate that our adopted temperatures should be correct to within $\pm 30\,$K except for a 
possible systematic zero-point shift common to all stars and amounting to maximally  100\,K.  In view of the
differential character of our study, such a shift will not affect our final conclusions. 

{\it Surface gravities:}
The \logg\ values were obtained from the $V$ magnitudes of the cluster stars \citep[from][]{Yadav&al:08},
by applying corrections for interstellar extinction, adequate bolometric corrections, and estimating the stellar 
masses from the isochrones of \citet{VandenBerg&al:07}. The errors in the differences \logg $-$\logg (M67-1194) are 
small and less than 0.03\,dex and mainly due to errors in the photometry and possibly varying extinction, as well as 
errors in the mass estimates from the isochrones. A significant source of much greater errors would be the possible 
binarity of some of the stars, leading to underestimates of surface gravity which may amount to 0.3\,dex. Such
an effect would be of particular interest since it could be selective -- it might affect our stars classified as 
TO and SG stars more than the MS stars since the positions of the binaries are expected to be raised 
in the colour-magnitude diagram (CMD). In the CMD for the cluster of \citet{Sandquist:04} we find most of the 
known binaries in such elevated positions, but none of our programme stars are among those, and they are also 
located low in the diagram. Thus, errors due to binarity are judged to only affect our results for few stars, 
and then with minor gravity (and abundance) effects. For the absolute values of \logg, additional errors enter 
due to errors in the value for M67-1194, estimated by \citep{Onehag&al:11} to be about 0.04\,dex. 
These errors are, however, compensated for in the abundance analysis by the differential approach chosen. 

The stars and their fundamental parameters are listed in Table~\ref{T1}. 

\begin{table}[htbp!]
  \caption{Fundamental parameters for the programme stars. The type of the star is indicated.
           MS: Main-sequence stars, TO: Turn-off point stars, SG: Subgiant-branch stars.}
  \centering
  \begin{tabular}[h]{rlccc}
    Star & Group & \teff\ \,[K] & log\,(g)\,[cms$^2$]  & \vmic \,[kms$^{-1}$]\\
    \hline \hline
    1194   & TWIN & 5780  & 4.44 &  1.0  \\
    \hline
    1116   & MS  & 6131  & 4.27 &  1.4  \\
    1221   & MS  & 6155  & 4.30 &  1.4  \\
    1265   & MS  & 6138  & 4.25 &  1.4  \\
    1367   & MS  & 6197  & 4.22 &  1.4  \\
    1568   & MS  & 6185  & 4.24 &  1.4  \\
    \hline
     963   & TO  & 6213  & 3.93 &  1.8  \\
    1629   & TO  & 6189  & 3.93 &  1.8  \\
    1783   & TO  & 6153  & 3.90 &  1.8  \\
    \hline
     863   & SG  & 6048  & 3.83 &  1.7  \\
    1188   & SG  & 6060  & 3.85 &  1.8  \\ 
    1248   & SG  & 6044  & 3.81 &  1.6  \\
    1258   & SG  & 6061  & 3.82 &  1.6  \\
    1320   & SG  & 6106  & 3.82 &  1.8  \\ 
   \hline
   \end{tabular} 
   \label{T1}
\end{table}

\subsection{Abundances}
The abundances were derived from the line strengths of a sample of absorption lines, selected to be generally on the 
linear part of the curve of growth, $\log\,(W_{eq}/\lambda) \le -5$, and to be relatively free of blends, as judged 
from the Kitt Peak Solar Atlas and the Vienna Atomic Line Data Base \citep[VALD,][]{Kupka&al:99}. Another important 
selection criterion was that lines used by \citet{Melendez&al:09} in their study of solar twins were included if 
possible. For spectral lines where blends were suspected to occur in one of the wings, the abundance estimates were 
based on the other half of the line. In practice, synthetic spectra were fitted to the useful parts of the observed 
lines and then the equivalent widths were calculated from the synthetic spectra of corresponding single lines 
that were next compared with model single lines for varying abundances. In a few cases, fits with synthetic spectra 
were used directly, not only for deriving abundances from severely blended but important lines but also when using 
relatively strong lines affected by hyper-fine structure, and for the Li\,I $\lambda 6707$ doublet. 
All lines used in the abundance analysis are listed and commented on in Table~\ref{T2} in the Appendix. 

The two O\,I lines in the sample, the high-excitation temperature sensitive $\lambda 6158$ and the blended forbidden
line $\lambda 6300$, are both rather weak in the temperature regime of the Sun and M67-1194. Chi-square minimised 
synthetic fits were calculated for both lines using SIU \citep{Reetz:91}, see also \citet{Onehag&al:11} for a more 
detailed description of the method and visualisation tool. Oscillator strengths for O\,I $\lambda 6300$ line and the  
Ni\,I blend were taken from \citet{Asplund&al:04} and \citet{Johansson&al:03}, respectively. The solar O\,I 
and Ni\,I abundances used in the calculations were taken from \citet{Asplund&al:04}.

\begin{figure*}
  \centering
   \resizebox{\hsize}{!}
    {\includegraphics[angle=90]{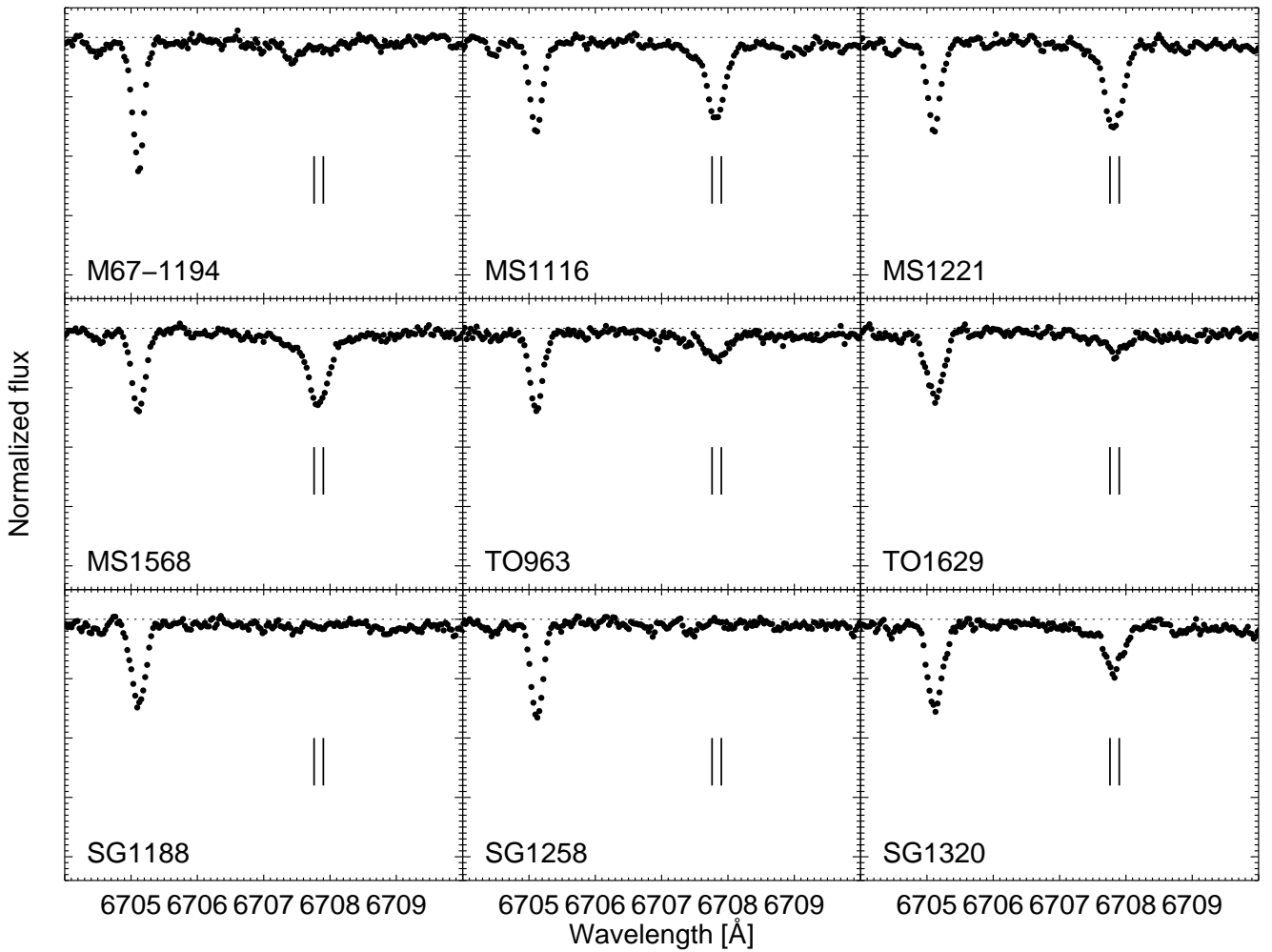}}
  \caption{Observed Li I $\lambda6707$ doublets (marked by two vertical lines) in 8 stars (type and designation given)
           selected to demonstrate the full range of Li line strengths. 
           The lines at $\lambda$6704.5 and $\lambda$6705.1 are Fe\,I lines.}           
  \label{F5}
\end{figure*}

The Li\,I doublet was found to vary considerably among the programme stars, and for four of them it was barely 
detectable at all. Some illustrative spectra are presented in Fig.~\ref{F5}. 

Abundances were derived for each spectral line individually relative to those of M67-1194, with the exception 
of Li for which a direct calibration on the solar Li doublet was applied. We used MARCS model atmospheres 
\citep{Gustafsson&al:08}. The models were calculated for each star with appropriate \teff\ and \logg\ and with solar 
abundances for all elements. The micro-turbulence parameter $\xi_{t}$ was varied until the individual abundances for 
Fe\,I lines with different strengths were as equal as possible for each star (see Table~\ref{T1}). The resulting 
abundances are presented in Table~\ref{T3}. Individual abundances for all lines are given in Table~\ref{T4}.

The errors in the abundances derived due to errors in the fundamental parameters are presented in Table~\ref{T5}. It is 
seen that the effects of errors in \teff\ are most significant, while effects due to errors in $\log g$ and in $\xi_t$ 
are negligible for most elements. Errors in effective temperatures are minimised by the differential approach while a 
temperature scale error of e.g. 100\,K would affect the absolute abundances considerably. The errors due to the inadequate 
assumptions of plane-parallel stratification, MLT convection, and LTE in the calculation of model atmospheres and spectra 
should also be minimised by the differential approach. However, they cannot be excluded from being of some significance, 
in particular for the differences in abundances obtained from the MS, TO, and SG stars, respectively (see below).

\begin{figure*}
  \centering
   \resizebox{\hsize}{!}
    {\includegraphics[angle=90]{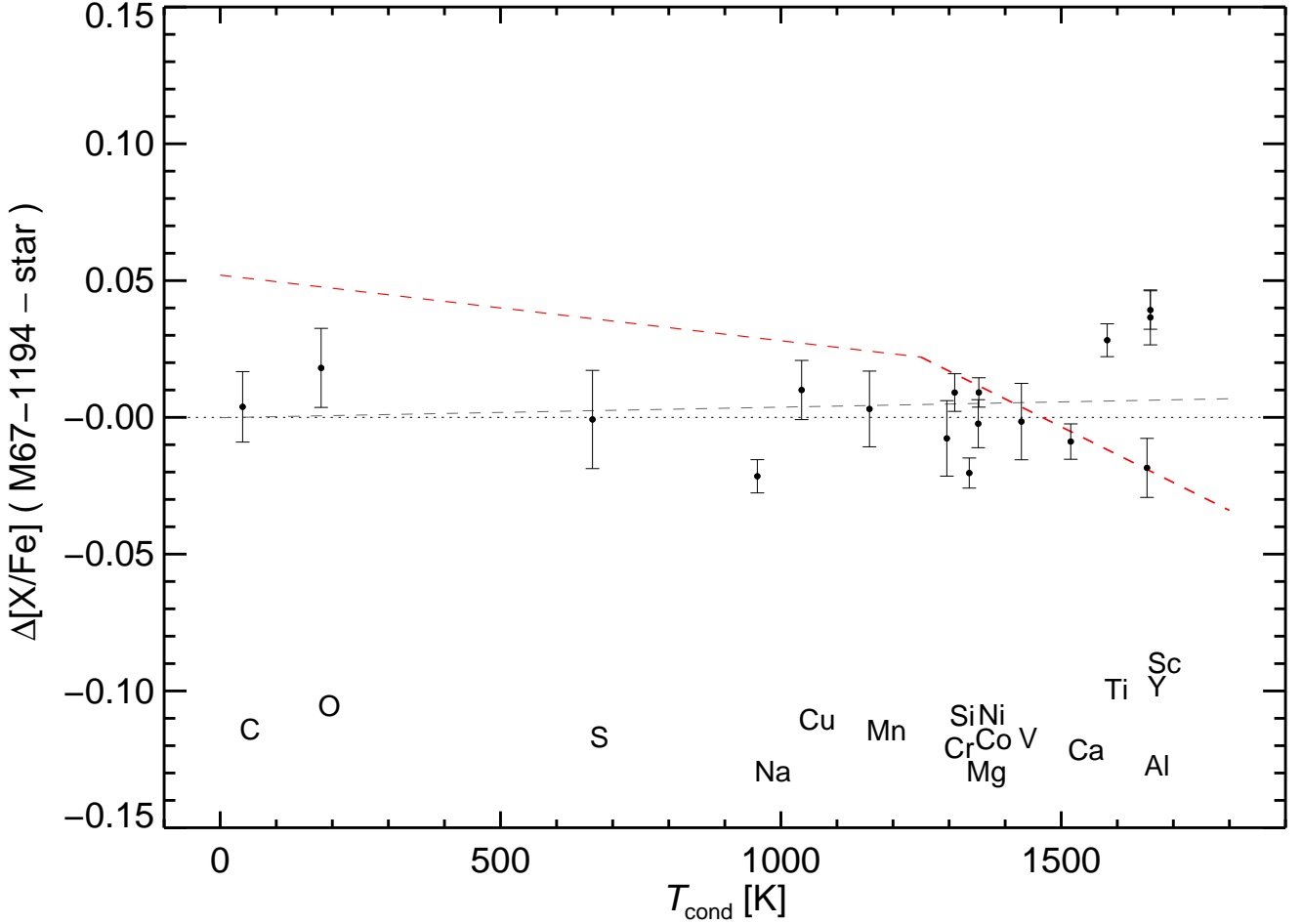}}
  \caption{The differences in logarithmic abundances between M67-1194 and the means for the other 13 programme 
           stars of the present study, plotted vs the condensation temperature of the elements \citep{Lodders:03}. 
           The red dashed line indicates the rough mean locus of the solar-solar twin abundances, according 
           to \citet{Melendez&al:09}. The grey dashed line is a regression line to the data.
           Error bars represent the RMS errors in the means for the entire sample ($\sigma/\sqrt N$).}
  \label{F6}
\end{figure*}

\section{Results and Discussion}
\label{sec:comp}
In the following, two different but interlinked aspects of the results will be discussed: (1) 
The overall cluster abundances where means of the chemical abundance of each element for all programme stars 
are at focus and (2) the systematic 
abundance differences between the three groups of stars, main-sequence stars (MS), turn-off point stars (TO), and 
subgiant-branch stars (SG). The first discussion is directly linked to the question of the abundance profile 
of the cluster, compared with that of the Sun and of Solar twins as explored by \citet{Melendez&al:09} and
\citet{Ramirez&al:09}, while the second issue is related to studies of diffusion by \citet{Korn&al:07} and 
later authors referred to in Sect.~\ref{sec:intro}. 
Finally, in the Conclusions (Sect.~\ref{sec:conclusions}) the interrelations between the conclusions drawn as regards 
the aspects (1) and (2) will be discussed.

\subsection{The abundance profile of the cluster and the dust-cleansing hypothesis}
\label{sec:dust}
\begin{figure}
  \centering
    \includegraphics[scale=1.1]{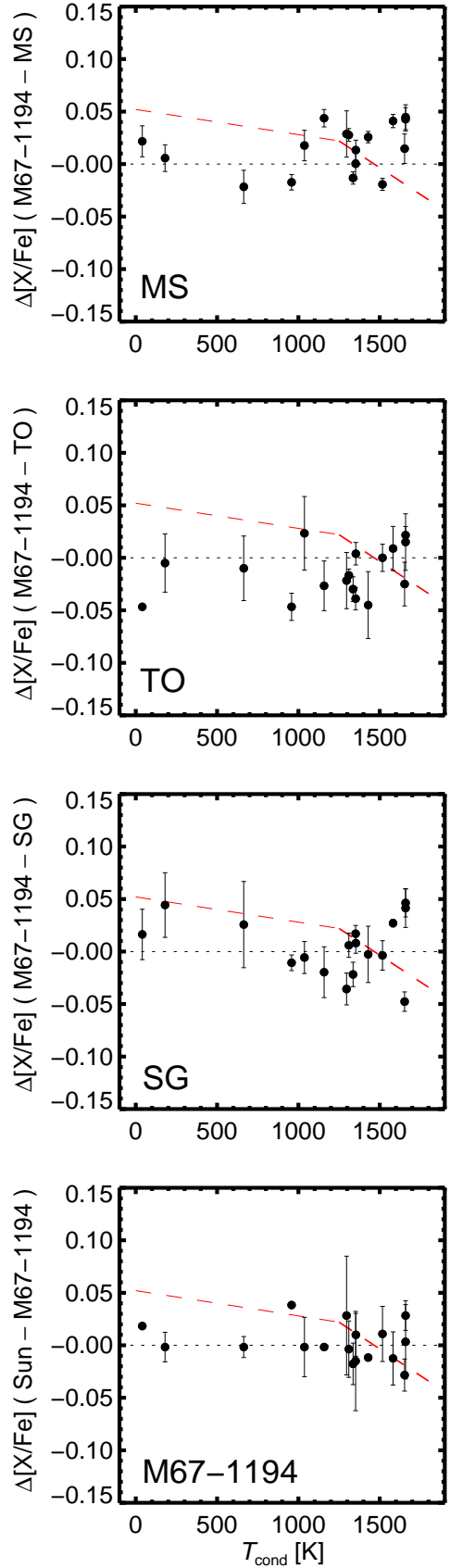}
  \caption{The abundance differences (M67-1194 $-$ group mean) where the means are taken for stars in the different groups, 
           Main-sequence stars, Turn-off point stars and Subgiant stars, respectively. The solar-twin line 
           of \citet{Melendez&al:09} is indicated (red dashed). The panel in the bottom displays 
           the differences Sun $-$ M67-1194. Error bars in the upper three panels represent the RMS errors 
           in the group means ($\sigma/\sqrt N$). In the bottom panel the error bars represent
           line-to-line scatter.}           
  \label{F7}
\end{figure}

As seen in Table~\ref{T3}  (see also Panel 4 of Fig.~\ref{F7}) the abundances derived here for M67-1194 are fully 
consistent with those of \citet{Onehag&al:11} which reinforces the conclusion from that paper (albeit based on 
the same observational data) that the star M67-1194 has an elemental composition very similar to that of the Sun.
We may thus take this star as a solar proxy with solar-like effective temperature and gravity, and closer 
to the Sun in chemical composition than the Solar twins explored by \citet{Melendez&al:09} and \citet{Ramirez&al:09}. 

In Fig.~\ref{F6} the mean abundances for the cluster stars relative to those of M67-1194 are plotted as a function of 
condensation temperature. Drawn in this figure is also the dashed red line showing the relation between abundance 
differences ($\Delta$[X(Fe] (Sun - Solar twins)) of \citet{Melendez&al:09}. Clearly, the observed abundances agree 
with the hypothesis that the abundances of the cluster stars in general match those of M67-1194 (and thus the Sun) 
better than those of the local twins. To put simple numbers on the certainty in this conclusion without referring to the 
somewhat uncertain estimated errors in the individual abundances, one may proceed as follows: We divide the full interval 
in condensation temperature $T_{\rm cond}$ into three sub-intervals: $T_{\rm cond}<1000$\,K, 1000\,K$ <$ $T_{\rm cond} < 1500$\,K,
and 1500\,K$ <$ $T_{\rm cond}$. The ratios of the numbers of points above and below the red dashed line in the diagram are, 
respectively 0/4, 0/8, and 4/1. Similarly, for the point distribution around the line corresponding to the composition of 
M67-1194 (the x-axis) the ratios are 2/2, 4/4, and 3/2. Assuming that the points were randomly distributed around the 
respective lines and that the error distributions are symmetrical around the true values, one finds by simple 
combinatorics that the probability for the composition to be equal to that of M67-1194 (i.e. this analysis to show the 
observed configuration of points around the x-axis in the figure) is three orders of magnitude greater than the 
probability that it would be equal to the composition of the solar twins in the field.

A more detailed study requires errors to be estimated in the various relative abundance means $<$[X/Fe]$>$ for the cluster 
stars. For the different elements we form these means for all our programme stars, and estimate the errors in the means 
as the standard deviation among the stars divided by $\sqrt N$, where $N$ is the number of stars. For elements where 
these errors are found to be less than 0.01 we adopt 0.01, in order not to underestimate the errors. We next 
hypothetically assume the true mean abundances of M67 to agree with the mean pattern of the Solar twins in the galactic field 
of \citet{Melendez&al:09}, i.e. to line up along the dashed line in Fig.~\ref{F6} when compared with M67-1194, our 
solar proxy, and that the apparent systematic difference between the observed values in the figure and the dashed 
relation is just circumstantial, due to the random errors in the analysis. When, using a Monte Carlo approach, we
apply a gaussian scatter characterised by the errors in the means, we find that less than 0.1$\%$ of all sets of 
such perturbed abundances fall closer to the abscissa than to the dashed line, i.e. with an absolute slope of less than 
$1.6\cdot 10^{-5}$\,dex/K in the figure. Thus, the hypothesis that the set of true mean abundances is closer to that of the 
Solar twins than to that of M67-1194 is several orders of magnitude less probable than the opposite hypothesis.
Indeed, even if the errors in $<$[X/Fe]$>$ were 
assumed to be as high as 0.04, a value which must be considered a severe overestimate, we find that the probability of 
the M67 stars to show the pattern of M67-1194 is 0.85, while the probability for them to show the pattern of the Solar 
twins is only 0.15.

A natural question is now to what extent this result is sensitive to the possible systematic errors. 
The most important of these errors for our present results is possibilities of errors in the temperature scale, 
which affect the abundances derived for volatile elements (low condensation temperatures) more than the 
refractories, as a result of the high excitation energy of the lower levels of C\,I and S\,I lines
used. An increase in the overall temperature scale by more than 200\,K is needed to bring the M67 stars 
anywhere near the abundance profile of the field dwarfs. In view of the discussion of the temperature scale 
in Sect.~\ref{sec:stellarparam}, we find such an error in the scale improbable. We thus conclude that the stars 
analysed in M67 have a mean abundance profile resembling that of the very solar-like star M67-1194, while
they deviate systematically from the mean results obtained for the Solar twins in the field by \citet{Melendez&al:09} 
and \citet{Ramirez&al:09}.

It is not clear, however, that this mean profile is representative of all M67 stars, as is displayed 
in Fig.~\ref{F7} where abundance patterns are shown for the different groups of stars. The subgiants 
do not seem to unambiguously show the solar likeness that the other stars demonstrate. If the results for all 
stars are scrutinised individually, at least 3 stars may show deviating tendencies, with patterns somewhat more similar 
to those of the Solar twins in the field. These stars are M67-863, 1320, and 1248. It is noteworthy that these 
deviating stars are all classified as subgiants. It is not appropriate to conclude that these subgiants
deviate because mixing of convective material from their interiors has erased departures at the surface, departures that 
could prevail for the MS and TO stars and could be due to diffusion in the stars or pollution of 
cleansed material from the proto-planetary disks. The convective zones of the subgiants at these relatively hot 
effective temperatures entail namely only about 2$\%$ of the stellar mass, a figure very similar to that of 
M67-1194, which is why dilution by convective mixing of a possible abnormal surface composition should not affect the
subgiants more than the solar-type dwarfs (we are indebted to Don VandenBerg for contributing these numbers).
Instead, one may speculate that the deviating abundances may be related to the higher mass of the subgiants 
(by typically $20\%$). However, the uncertainties in the abundances, and the small sample of subgiants, do 
not admit us to draw any certain conclusions in those directions. 

We note in passing that the SG star 1320 shows some anomalously small abundances from the OI $\lambda$6158,
SI $\lambda$6052, and FeII $\lambda$5454 lines, which might suggest a lower effective temperature than we have 
adopted. However, we do not find any evidence for a lower temperature from our different temperature estimates.
It could also be suspected that the stellar spectrum is contaminated by a companion; however, we have 
not found any further evidence for this. If 1320 would be excluded from the sample, the subgiants would show a 
more solar-similar pattern in Fig.~\ref{F7}.

A possible explanation for our result in Fig.~\ref{F6} is that the gas in dense proto-clusters might have 
been cleansed from dust by hot stars early on in the evolution of the cluster. \citet{Draine:11} has explored the 
possibilities for dust to drift out of HII regions 
as a result of the stellar radiation pressure on the grains.  For his static equilibrium models of HII regions 
he finds self-similar solutions, dependent on three parameters that characterise the stellar radiation, the gas 
to dust ratio and the product $Q_0 \times n_{rms}$, respectively. Here $Q_0 $ is the number of photons that may 
ionise hydrogen emitted per second and $n_{rms}$ is the rms density of protons within  the ionised region. Adopting 
the estimate by \citet{Hurley&al:05} that M67 early on contained nearly 19000 solar masses in stars, one finds 
that the original Giant Molecular Cloud (GMC) from which the cluster formed had a mass around $10^5$ M$_\odot$ if the 
time-integrated star formation efficiency $\epsilon$ was as high as 20\%, a number estimated for GMCs in the 
Milky Way \citep{Murray:11}. With more traditional lower values of $\epsilon$ the mass would have been even 
greater. Adopting a standard stellar mass distribution function \citep{Cox:00}, we then find 
that about 1 star formed in the cluster should have had a mass above 25 solar masses. For this case 
(corresponding to $Q_0 \sim 10^{49}$ and $n_{rms} \sim 500$ cm$^{-3}$) we find drift speeds of 0.1\,kms$^{-1}$ or more, 
and a characteristic time for the dust grains to drift out of the model HII region of 2\,Myr, from Figs. 8 and 9 of 
\citet{Draine:11}. The latter value is fairly independent of the parameters of the model, although, as shown by 
\citet{Draine:11} magnetic fields in the gas may increase the time by orders of magnitude. A higher total gas 
mass of the proto-cluster would lower the drift time in inverse proportion.

For the suggested dust cleansing of the proto-cluster cloud by hot stellar radiation to be the explanation for 
the composition found for the M67 stars, as compared with the solar twins in the galactic field which would then 
presumably have formed in less massive clusters where no high-mass star was formed, one must also assume a 
star-formation history in the proto-cluster cloud to occur in several generations, where the radiating massive 
star(s) formed before most of the lower-mass cluster stars. This could then be an example of a triggered star 
formation scenario, as discussed by \citet{Sharma&al:07}, \citet{Jose&al:08}, and \citet{Pandey&al:08,Pandey&al:13},
on the basis of studies of stellar distributions in OB associations. One could speculate that the sign of 
deviating abundances for the subgiants with their somewhat higher masses could be the result of an earlier star 
formation than for the solar-type dwarfs, at epochs when the cleansing had still not been efficient.  

However, it is unclear whether this mechanism of dust cleansing at all gives the result observed. Indeed, as was pointed 
out by \citet{Draine:11} the true HII regions are not static but their ionisation fronts are expanding with 
speeds that may well supersede the dust drift speeds. If the "collect and collapse process" of 
\citet{Elmegreen&Lada:77} were at play, where the neutral gas surrounding the HII region is collected between 
the ionisation front and its preceding shock front, and next collapses gravitationally, triggered by the pressure 
of the ionised gas (cf. simulations by \citealt{Hosokawa&Inutsuka:05, Hosokawa&Inutsuka:06} and \citealt{Dale&al:07}), 
the dust drift may also not be swift enough to deplete the dust in the swept-up gas. It remains to explore 
whether the radiation-driven implosion (RDI) of molecular cloud condensations in the proto-cluster 
\citep[][and references therein]{Miao&al:06} could admit dust depletion to take place.

The proposal that dust-cleansing is the explanation for the composition of M67 raises another question. We 
have found that the different stars in M67 suggest a depletion of refractories by at the most about 20\%. 
Certainly, this does not correspond to a total depletion of all dust which once existed in the proto-cluster 
cloud, if the degrees of condensation typical of interstellar dust would apply. The latter depletions are often 
found to be one order of magnitude greater \citep{Savage&Sembach:96}. 
In this case, an efficient evaporation and mixing of the gas must have taken place before the programme stars were formed.

\subsection{The effects of diffusion in the stars and the lithium abundances}

\begin{figure}
  \centering
   \resizebox{\hsize}{!}
    {\includegraphics[angle=90]{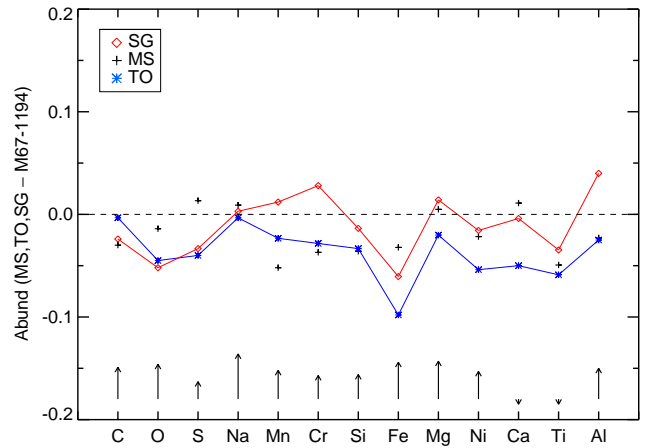}}
  \caption{The differences in logarithmic abundances between the three stellar groups and M67-1194
           for the chemical elements analysed. A systematic difference between the SG and TO stars
           is apparent and the mean value for the 
           MS stars falls in between the other two groups. Arrows indicate model predictions
           by O.~Richard (priv.comm.) of the effects of atomic diffusion (with gravitational settling and
           radiative levitation) on the surface abundances when moving from TO stars to SG stars in 
           the temperature regime covered in this study.}           
  \label{F8}
\end{figure}

In our analysis we attempted to trace another effect which is of interest to studies of stellar 
structure and evolution beyond classical quasi-hydrodynamic models.
If the observed [X/H] values, i.e. abundances relative to M67-1194, of all elements except Li are 
averaged for all stars one finds the value $-0.017\,\pm\,0.003$ (the error being the mean error in 
the mean), while the corresponding predicted value from the model calculations of O.~Richard 
(private communication) with atomic diffusion (but without additional mixing) is $-0.028\,\pm\,0.002$.
Here, his models for an age of M67 of 3.7\,Gyr were adopted. 
These values are calculated relative to our standard star M67-1194 for which [X/H] is set to zero;
this star is cooler and has therefore a considerably deeper convection zone than 
most of our programme stars which is why the diffusion effects are expected to be smaller. The corresponding 
observed mean for the TO and MS stars, i.e. with the five SG stars with their deeper convection zones excluded, 
is, as one would expect if diffusion is active, absolutely larger, $-0.026\pm0.004$ which is compatible 
with the model predictions. A differential study of TO and MS stars relative to SG stars
gives an observed difference in the means of $-0.021 \pm 0.007$ which again is compatible with the
predictions. Thus, these different tests seem to suggest that diffusion is active. However, the diffusion 
models also predict selective effects when different elements are compared. Among the elements studied here,
one would thus expect diminishing abundances of Ca and Ti by about 0.01\,dex when the stars proceed from 
the TO to the SG, a behaviour in contrast to the increase in atmospheric abundances shown by the other 
elements. This effect is not demonstrated by our data which instead indicate a similar behaviour 
for Ca and Ti as for, e.g., Fe. 

These effects are illustrated for the individual elements in Fig.~\ref{F8} where also the effects of
atomic diffusion according to the model calculations by Richard are indicated by arrows. 
Corrections of the temperatures for the groups relative to one another might 
possibly explain the abundance differences seen in the figure: a possible increase in \teff\ by 50\,K 
for the TO stars relative to the SG stars, as may be suggested by the ionisation equilibria of Fe and Ti (cf. 
Fig.~\ref{F4}), would explain the effects for most of the temperature-sensitive species 
(Mg\,I, Ca\,I, Ti\,I, Fe\,I etc.).

We conclude that our results suggest the presence of diffusion-induced heavy-element 
abundance differences in solar-metallicity stars. The correspondence between theory and observation is acceptable,
albeit not equally so for all elements. Some precaution must be exercised, since systematic errors in even differential 
chemical analyses of stars are still noticeable. As one example, brought forward by the referee, one may take the 
case of the binary 16 Cyg A and B, two solar analogues which have recently been analysed on the basis of spectra of 
very high quality, independently by \citet{Ramirez&al:11} and \citet{Schuler&al:11}. While the former found an 
overabundance of all heavy elements of the A component by 0.04\,dex as compared to the B component, the latter group 
obtained more similar or identical abundances for the two components. This illustrates the fact that even accurate
analyses, based on first-class data, may suffer from systematic errors that are significant when accuracies better
than 0.05\,dex are needed. Therefore, the differences traced by us for the subgroups of the M67 stars must still 
be regarded as somewhat preliminary. Independent verification with still better data is important.

We note that according to the model predictions, the metallicities of all observed stars have to be corrected
upwards if one is interested in the original composition of M67. Judging from the Richard models, its
likely value lies close to +0.06 (cf. the Sun's original composition of +0.05, \citealt{Lodders:10}).
Such an increase in the interior metallicity of stars in M67 also improves a fit of the CMD at the TO
region by producing a gap representing the occurrence of sustained core convection on the main sequence
(see the discussion of \citealt{VandenBerg&al:07}).

\begin{figure}
  \centering
   \resizebox{\hsize}{!}
    {\includegraphics[angle=90]{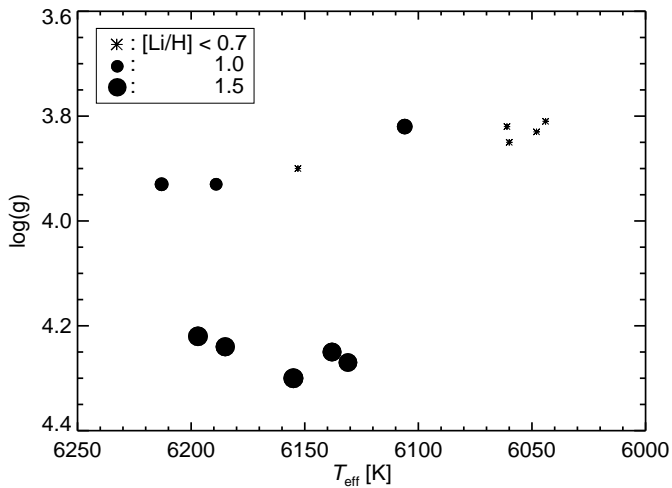}}
  \caption{The stars plotted in the \teff\ - \logg\ plane, with symbol sizes
           indicating the Li abundances. Asterisks show upper abundance limits.}
  \label{F9}
\end{figure}

Lithium may be used as a sensitive tracer of the mixing history in the outer layers of solar-type stars. 
In Fig.~\ref{F9} we have plotted our Li abundance estimates as different symbol sizes in a \teff\ -\logg\
diagram. Our results are fairly consistent with those of \citet{Pasquini&al:08} who measured Li abundances 
for a great number of M67 stars, although their results, based on more spectra of lower quality, show a 
greater scatter in abundances for stars in the parameter space of our MS stars. The errors in our abundances 
are typically about 0.1\,dex, or greater for the stars with lowest abundances due to the weakness of the 
Li\,I $\lambda$6707 doublet (here our values only represent upper limits, as indicated in the figure). 
It is seen that the abundances for the MS stars are typically 1.5\,dex higher, with only a small scatter, 
than the solar photospheric abundance, which may be compared with the difference between the solar system 
logarithmic meteoritic value and the photospheric value of 2.2, as listed by \citet{Asplund&al:09}. 
Note that our [Li/H] values given in tables, figures and in the text above are all logarithmic values, 
determined directly relative to the Sun, i.e. unlike for the other abundances M67-1194 has not been used for 
Li as a solar proxy. A correlation between stellar mass and Li abundance has been found in studies of M67 
\citep{Pace&al:12, Li&al:12}. The higher Li abundance determined for our MS stars are, according 
to these studies, likely reflecting the higher mass for the MS stars compared to 
the Sun. For the TO stars there is a significant spread towards lower values. Among the five SG stars, we find a 
very large scatter, 1.2\,dex, in lithium abundances. Star 1320, discussed in Sect.~\ref{sec:dust}, 
seems unmixed while the four other 
SG stars retain little if any of their original lithium. This mixing would be $non-standard$ in nature and 
may be related to the physical processes giving rise to the lithium dip seen in younger open clusters. 
Lithium itself may obviously serve to constrain the mixing scenario. 

\section{Conclusions}
\label{sec:conclusions}
Our main result is the following: a dominant fraction of the stars in M67 show abundance patterns more similar 
to the Sun than most solar twins in the Solar neighbourhood do. This supports the conclusion that the 
Sun was formed in a dense stellar environment. A probable but not certain reason for this specific 
pattern is the action of dust cleansing in the primordial states of the cluster, as well as in the 
proto-stellar cloud that formed the Sun. 

Even if there are several independent indications that the Sun was once formed in a dense environment
it remains to be explained why its degree of depletion of refractories agrees quantitatively with that 
of the M67 stars. A tempting, though not unproblematic, solution of this problem would be that the Sun 
originated in M67. In order for this to be reasonable, one must explain why the Sun now has an 
orbit close to the Galactic plane, while M67 is presently high above the plane at a $z$ distance of 450\,pc. 
As shown by \citet{Pichardo&al:12}, the Sun could hardly have elapsed from the cluster into its present 
orbit without loosing significant parts of its planetary system. It remains to be seen if the converse 
scenario, where the cluster was formed in an orbit close to the plane and lost the Sun before it was 
diverted into a higher-altitude orbit, e.g. as a result of a collision with a Giant Molecular Cloud or 
through the effect of Galactic spiral arms is viable.

Recently \citet{Melendez&al:12} studied the solar twin HD101364 (HIP56948) and found it to have a 
composition very close to solar, thus departing significantly from the composition of most twins in the 
solar neighbourhood. The authors estimated the age of HD101364 to 3.52\,$\pm$\,0.68\,Gyrs, which is 
compatible with the age estimates for M67. There might be a possibility that this star once formed in the 
cluster -- the typical distance in between the almost 20,000 stars that have left it \citep{Hurley&al:05} 
since it formed  can be estimated to be about 100\,pc, provided that they are distributed like Pop I stars 
in the Galactic disk at a distance 7\,kpc - 9\,kpc from the Galactic centre. This would make the chances of 
finding a star brighter than 9th magnitude ($V$(HD101364)=8.67) non-negligible. We note, however, that the 
velocity perpendicularly to the Galactic plane of this star is only $W$\,=\,$-$5.93\,kms$^{-1}$ which does 
not admit it to reach the height above the plane that M67 has today, while its $U$ velocity is high, 
36\,kms$^{-1}$. Anyhow, this interesting star is valuable as a template for comparative studies of 
solar twins in M67 and other clusters.

Several alternative explanations for the systematic abundance difference between the Sun and the field 
twins were discussed by \citet{Melendez&al:09} and \citet{Gustafsson&al:10}. An interesting 
possibility is that some stars, like the Sun, only had a shallow convective zone when the gas of the 
proto-planetary disk depleted by planetary formation was dumped onto its surface, and never were fully mixed 
after that. Such an unconventional solar formation history could result from rapid episodic accretion onto 
a small stellar core, as has been suggested by \citet{Baraffe&Chabrier:10}. Our M67 project was partly 
devised as a test of this possibility -- an outcome proving that stars in a dense environment all showed 
the solar abundance profile would then suggest the dust-cleansing hypothesis to be the right one and 
disprove the local gas-disk pollution hypothesis. True enough, the result of the present study supports 
the first hypothesis. However, it is still possible that the local-disk pollution hypothesis is valid. 
One might for instance argue that the proto-stellar disks of stars in dense environments are affected by 
radiation, or even gravitational disturbances, from nearby stars, such that the instabilities in the disk 
are accentuated, promoting rapid episodic accretion onto the star.  

Our tentatitve detection of systematic heavy-element abundance differences among stars in a solar-metallicity 
cluster opens up the possibility of observationally constraining (and subsequently correcting for) the amount 
of atomic diffusion affecting the surface abundances of late-type stars like the Sun. It remains 
to combine this finding with the Li depletion data into a coherent picture.   

In the discussion above we have tried to separate the two phenomena observed: (1) the similarity between 
the overall abundance profile of M67 and that of the Sun, being different from the abundance profile of 
the twins in the solar neighbourhood and (2) the indications that the atmospheres of TO and MS stars seem to be 
depleted in heavy elements as compared to those of the SG stars. However, we have also pointed out the fact 
that these two phenomena are interrelated, since the composition of the SG stars according to diffusion 
models are presumably less affected by diffusion and thus should show a composition closer to the cloud 
from which the cluster once formed. As we saw in Fig.~\ref{F7} there is a tendency, yet with the present data 
of marginal significance, for the SG stars not to clearly show the composition profile of the Sun. Instead 
they could be more similar to the twins in the solar neighbourhood. Now, if the SG stars are considered to be 
essentially free of the diffusion effects, could one not argue that what we see in the TO and MS stars 
is only reflecting {\it one} phenomenon: the effects of diffusion (or other element separation) intrinsic 
to the stars, and that the original abundance profile indeed was quite similar to that of the field twins? 
The main problem with that argument is then why the diffusion effects found in M67 are not shown by 
the field solar twins except for the Sun itself. One could speculate that, e.g., the rotation of the cluster 
stars, or their primordial magnetic fields, or their accretion history, were systematically different from 
that of the field stars (but not of the Sun), leading to different and more efficient element separation in 
these stars than for the field stars. Such hypotheses have, however, little support theoretically or 
observationally, and could as well go the other way. If alternatively the dust cleansing scenario is 
advocated to explain the abundance profile of the TO and MS stars, one must explain why it does not seem to 
have affected the SG stars as much. The hypothesis that the SG stars, due to their systematically 
higher masses, were formed before the cloud was fully cleansed or that their proto-planetary disks evolved 
differently as compared to the lower-mass TO and MS stars, are also rather {\it ad hoc}.  In conclusion 
it seems clear that a study at higher resolution and S/N, and with more cluster stars, is needed to settle 
these issues.   

Our two findings, indicating effects of fractionation of chemical elements in star-forming regions 
or even within the individual solar-mass stars, have important consequences for studies where solar-type 
stars are used to explore the chemical evolution of the Galaxy. Obviously, it cannot be taken for granted 
that solar-type stars, in spite of their comparatively well-mixed surface layers, fully reflect the chemical 
composition of their primordial gas and dust clouds. 

\acknowledgement{
Several colleagues have contributed importantly to the present paper. 
Eric Stempels devised the method to correct the observed spectra for light-contamination in the 
fibre bundles of FLAMES-UVES. Frank Grundahl supplied photometric data and fundamental-parameter 
estimates for the programme stars. Olivier Richard allowed us to use results from calculations of 
atomic-diffusion effects in models of stellar evolution representing M67. Kjell Eriksson calculated 
effects of \logg\ and microturbulence variations on the $V-K_{\rm s}$ and $V-I_{\rm c}$ colours and 
Don VandenBerg provided data on evolutionary models of solar-mass stars. We also thank the anonymous
referee for valuable comments on the manuscript.

\bibliographystyle{aa}
\bibliography{22663ref}

\begin{appendix}

\section{Tables of lines, derived abundances and error estimates}

   \begin{table}[htbp]
  \caption{Lines used in the abundance analysis. Column description (left to right):
           Atomic element, wavelength, excitation energy, oscillator strength according to VALD, 
           measured equivalent width from the Kitt Peak Solar Atlas, comments.}
  \begin{tabular}[h]{lccccl}
  Elem. & $\lambda$\,[\AA] & $\chi$\,[eV] & $\log gf$ & W$_{\rm eq\,\, \odot}$ [m\AA] & \\
  \hline \hline
  Li\,{\sc I}  & 6707.761 & 0.00 &    0.167 &    4.1 & $synth$    \\ 
  C\,{\sc I}   & 5380.337 & 7.68 & $-$1.842 &   23.8 & $b$  \\
  O\,{\sc I}   & 6158.186 & 10.74& $-$0.409 &   4.8  & $siu$      \\
  O\,{\sc I}   & 6300.304 & 0.00 & $-$9.819 &   6.2  & $siu$      \\
  Na\,{\sc I}  & 6154.226 & 2.10 & $-$1.560 &   40.6 & $b$  \\
  Na\,{\sc I}  & 6160.747 & 2.10 & $-$1.260 &   62.2 & $b$  \\
  Mg\,{\sc I}  & 5711.088 & 4.35 & $-$1.833 &  108.0 & $s$        \\
  Mg\,{\sc I}  & 5183.604 & 2.72 & $-$0.180 &        & $s,synth$  \\
  Al\,{\sc I}  & 6696.185 & 4.02 & $-$1.576 &   39.4 & $b$  \\
  Al\,{\sc I}  & 6698.673 & 3.14 & $-$1.647 &   22.4 & $b$  \\
  Si\,{\sc I } & 5665.555 & 4.92 & $-$2.040 &   42.9 &            \\
  Si\,{\sc I } & 5690.425 & 4.93 & $-$1.870 &   53.3 &            \\
  Si\,{\sc I } & 6125.021 & 5.61 & $-$1.464 &   35.2 &            \\
  Si\,{\sc I } & 6131.852 & 5.62 & $-$1.615 &   26.0 & $b$  \\
  Si\,{\sc I } & 6721.848 & 5.86 & $-$1.516 &   49.7 &            \\
  S\,{\sc I}   & 6046.027 & 7.87 & $-$1.030 &   11.6 & $b$  \\
  S\,{\sc I}   & 6052.674 & 7.87 & $-$0.740 &   13.4 &            \\
  S\,{\sc I}   & 6757.171 & 7.87 & $-$0.310 &   15.6 & $b$  \\ 
  Ca\,{\sc I}  & 5590.114 & 2.52 & $-$0.571 &   96.2 & $b,s$\\
  Ca\,{\sc I}  & 5867.562 & 2.93 & $-$1.570 &   27.9 &            \\
  Ca\,{\sc I}  & 6166.439 & 2.52 & $-$1.142 &   73.8 & $s$        \\
  Ca\,{\sc I}  & 6169.042 & 2.52 & $-$0.797 &  100.2 & $b,s$\\
  Sc\,{\sc II} & 5684.202 & 1.51 & $-$1.074 &   39.4 & $b$  \\
  Sc\,{\sc II} & 6245.637 & 1.51 & $-$1.030 &   36.6 & $b$  \\
  Ti\,{\sc I } & 5219.702 & 0.02 & $-$2.292 &   30.2 & $b$  \\
  Ti\,{\sc I } & 6126.216 & 1.07 & $-$1.425 &   24.1 &            \\
  Ti\,{\sc I } & 6258.102 & 1.44 & $-$0.355 &   52.9 &            \\
  Ti\,{\sc II} & 5490.690 & 1.57 & $-$2.430 &   23.6 &            \\
  Ti\,{\sc II} & 6491.561 & 2.06 & $-$1.793 &   41.4 & $b$  \\
  V\,{\sc I }  & 5670.853 & 1.08 & $-$0.420 &   20.3 &            \\
  V\,{\sc I }  & 6090.214 & 1.08 & $-$0.062 &   35.7 &            \\
  Cr\,{\sc I}  & 5238.961 & 2.71 & $-$1.305 &   18.2 & $b$  \\
  Cr\,{\sc I}  & 5287.178 & 3.44 & $-$0.907 &   12.4 & $b$  \\
  Mn\,{\sc I}  & 5377.637 & 3.84 & $-$0.109 &   50.3 &            \\
  Fe\,{\sc I}  & 5295.312 & 4.41 & $-$0.967 &   91.4 & $b$  \\
  Fe\,{\sc I}  & 5522.446 & 4.21 & $-$1.514 &   65.2 &            \\
  Fe\,{\sc I}  & 6151.617 & 2.18 & $-$1.530 &  125.4 &            \\
  Fe\,{\sc I}  & 6229.226 & 2.85 & $-$2.482 &   84.7 &            \\
  Fe\,{\sc I}  & 6498.938 & 0.96 & $-$2.027 &  108.6 & $b$  \\
  Fe\,{\sc II} & 5414.073 & 3.22 & $-$3.645 &   32.0 & $b$  \\
  Co\,{\sc I}  & 5342.695 & 4.02 &    0.690 &   35.4 & $b$  \\
  Co\,{\sc I}  & 5530.774 & 1.71 & $-$2.060 &   17.2 & $b$  \\
  Co\,{\sc I}  & 5647.234 & 2.28 & $-$1.560 &   14.8 &            \\
  Ni\,{\sc I}  & 5589.357 & 3.90 & $-$1.140 &   31.2 & $b$  \\
  Ni\,{\sc I}  & 6086.276 & 4.27 & $-$0.530 &   46.2 & $b$  \\
  Ni\,{\sc I}  & 6130.130 & 4.27 & $-$0.960 &   23.4 &            \\
  Ni\,{\sc I}  & 6204.600 & 4.09 & $-$1.100 &   23.2 &            \\
  Ni\,{\sc I}  & 6223.981 & 4.11 & $-$0.910 &   29.6 &            \\
  Ni\,{\sc I}  & 6772.313 & 3.66 & $-$0.980 &   52.7 &            \\
  Cu\,{\sc I}  & 5218.197 & 3.82 &    0.476 &   58.0 & $b,s$\\
  Cu\,{\sc I}  & 5220.066 & 3.82 & $-$0.448 &   15.2 & $b$  \\
  Y\,{\sc II}  & 4900.120 & 1.03 & $-$0.090 &   59.2 & $b,s$\\
  Y\,{\sc II}  & 5087.416 & 1.08 & $-$0.170 &   50.6 & $b,s$\\
  \hline
   \end{tabular} 
   \tablefoot{$b$: blend in wing(s), parts of line used to calculate W$_{\rm eq}$; 
              $s$: strong- or intermediate strong line in stellar spectra, $\log\,(W_{eq}/\lambda) > -5 $;
              $synth$: abundance determined directly from synthetic fit; 
              $siu$: abundance determination and synthetic fit made in SIU.}
  \label{T2}
\end{table}

\begin{onecolumn}
\begin{landscape}

\begin{table*}[htbp]
  \caption{Resulting abundances for the programme stars. Main sequence (MS), turn-off point (TO), and subgiant-branch (SG) 
           stars are all presented relative to M67-1194, log[abund(star)]$-$log[abund(M67-1194)], except for Li that is presented relative to the Sun. 
           The rightmost column displays the abundance difference between M67-1194 and the (Kitt Peak) Sun, log[abund(M67-1194)]$-$log[abund(Sun)].}
  \begin{tabular}[h]{lrrrrrrrrrrrrrr}
Elem   & MS1116  &  MS1221  & MS1265  &  MS1367 & MS1568 &  TO963  &  TO1629 & TO1783  &  SG863  & SG1188  &  SG1248 & SG1258  &   SG1320 & M67-1194 \\
 \hline \hline                                                                                                                        
 Li\,I &    1.50 &     1.60 &    1.52 &    1.58 &    1.54&    1.11 &    1.02 &    0.68 &    0.41 &    0.42 &    0.42 &    0.08 &     1.26 &    0.11 \\
  C\,I & $-$0.07 &     0.02 & $-$0.06 &    0.00 & $-$0.04& $-$0.02 &    0.03 & $-$0.02 & $-$0.04 & $-$0.04 &    0.02 & $-$0.01 &  $-$0.05 & $-$0.01 \\
  O\,I & $-$0.02 &  $-$0.04 & $-$0.03 &    0.01 &    0.01& $-$0.08 & $-$0.04 & $-$0.01 &    0.01 & $-$0.04 & $-$0.04 & $-$0.04 &  $-$0.15 &    0.01 \\
 Na\,I &    0.00 &     0.01 & $-$0.01 &    0.07 & $-$0.02& $-$0.04 &    0.06 & $-$0.03 &    0.04 & $-$0.05 &    0.04 & $-$0.01 &     0.01 & $-$0.03 \\
 Mg\,I & $-$0.01 &  $-$0.01 & $-$0.02 &    0.07 & $-$0.01& $-$0.01 &    0.00 & $-$0.05 &    0.06 & $-$0.08 &    0.02 &    0.01 &     0.06 &    0.03 \\
 Al\,I & $-$0.04 &  $-$0.02 & $-$0.01 &    0.00 & $-$0.05&    0.00 & $-$0.03 & $-$0.05 &    0.09 &    0.00 &    0.06 &    0.01 &     0.06 & $-$0.02 \\
 Si\,I & $-$0.06 &  $-$0.03 & $-$0.07 &    0.01 & $-$0.03& $-$0.04 & $-$0.01 & $-$0.05 &    0.03 & $-$0.04 &    0.00 & $-$0.03 &  $-$0.03 &    0.01 \\
  S\,I &    0.03 &     0.05 & $-$0.03 &    0.02 & $-$0.01& $-$0.11 & $-$0.01 & $-$0.01 & $-$0.05 &    0.05 & $-$0.06 & $-$0.05 &  $-$0.05 &    0.01 \\
 Ca\,I & $-$0.01 &     0.01 & $-$0.02 &    0.08 &    0.00& $-$0.04 & $-$0.03 & $-$0.09 &    0.04 & $-$0.05 &    0.00 & $-$0.04 &     0.04 &    0.00 \\
Sc\,II & $-$0.09 &  $-$0.02 & $-$0.07 & $-$0.01 & $-$0.07& $-$0.07 & $-$0.04 & $-$0.10 & $-$0.01 & $-$0.09 &    0.00 & $-$0.09 &  $-$0.08 & $-$0.01 \\
  Ti   & $-$0.06 &  $-$0.06 & $-$0.07 &    0.00 & $-$0.05& $-$0.03 & $-$0.05 & $-$0.09 &    0.02 & $-$0.10 &    0.00 & $-$0.04 &  $-$0.02 &    0.02 \\
  V\,I & $-$0.03 &  $-$0.04 & $-$0.09 &    0.01 & $-$0.03&    0.03 &    0.04 & $-$0.09 &    0.01 & $-$0.13 &    0.06 & $-$0.04 &     0.09 &    0.02 \\
 Cr\,I & $-$0.04 &  $-$0.07 &    0.00 & $-$0.03 & $-$0.04&    0.00 & $-$0.04 & $-$0.04 &    0.12 & $-$0.04 &    0.03 & $-$0.03 &     0.07 & $-$0.02 \\
 Mn\,I & $-$0.08 &  $-$0.05 & $-$0.11 &    0.03 & $-$0.05& $-$0.02 &    0.04 & $-$0.09 &    0.02 & $-$0.08 &    0.13 & $-$0.02 &     0.01 &    0.01 \\
  Fe   & $-$0.02 &  $-$0.01 & $-$0.05 &    0.05 & $-$0.01& $-$0.06 & $-$0.02 & $-$0.07 &    0.06 & $-$0.08 &    0.02 & $-$0.03 &  $-$0.01 &    0.01 \\
 Co\,I & $-$0.05 &     0.00 & $-$0.04 &    0.09 & $-$0.04&    0.00 &    0.01 & $-$0.04 &    0.02 & $-$0.11 &    0.01 & $-$0.03 &     0.00 &    0.02 \\
 Ni\,I & $-$0.04 &  $-$0.01 & $-$0.03 &    0.02 & $-$0.04& $-$0.05 & $-$0.01 & $-$0.10 &    0.03 & $-$0.06 &    0.03 & $-$0.04 &  $-$0.03 &    0.00 \\
 Cu\,I & $-$0.03 &  $-$0.02 & $-$0.04 &    0.04 & $-$0.09& $-$0.06 & $-$0.11 & $-$0.06 &    0.09 & $-$0.12 & $-$0.01 &    0.02 &     0.01 &    0.01 \\
 Y\,II & $-$0.09 &  $-$0.06 & $-$0.09 & $-$0.02 & $-$0.01& $-$0.03 & $-$0.07 & $-$0.10 & $-$0.01 & $-$0.12 &    0.04 & $-$0.07 &  $-$0.09 & $-$0.01 \\
 \hline
  \end{tabular}
  \label{T3}
  \end{table*} 

\end{landscape}
\end{onecolumn}

\begin{onecolumn}
\begin{landscape}

\begin{table*}[htbp]
  \begin{tiny}
  \caption{Derived abundances for all individual lines used in the analyses. For further explanations see table head of 
           Table \ref{T3}.}
  \begin{tabular}[h]{lrrrrrrrrrrrrrrrr}
Elem   & Wavl.[\AA]& MS1116  & MS1221  & MS1265  & MS1367  & MS1568  & TO963   & TO1629  & TO1783  &  SG863  & SG1188  & SG1248  & SG1258  & SG1320  & M67-1194\\
\hline \hline                                                                                                                                                            
  Li\,I & 6707.761 &    1.50 &    1.60 &    1.52 &    1.58 &    1.54 &    1.11 &    1.02 &    0.68 &    0.41 &    0.42 &    0.42 &    0.08 &    1.26 &    0.11 \\
   C\,I & 5380.337 & $-$0.07 &    0.02 & $-$0.06 &    0.00 & $-$0.04 & $-$0.02 &    0.03 & $-$0.02 & $-$0.04 & $-$0.04 &    0.02 & $-$0.01 & $-$0.05 & $-$0.01 \\
   O\,I & 6158.186 & $-$0.15 & $-$0.15 & $-$0.13 & $-$0.10 & $-$0.10 & $-$0.14 & $-$0.11 & $-$0.12 & $-$0.06 & $-$0.09 & $-$0.13 & $-$0.09 & $-$0.24 &    0.00 \\
   O\,I & 6300.304 &    0.11 &    0.08 &    0.07 &    0.11 &    0.12 & $-$0.02 &    0.02 &    0.10 &    0.09 &    0.01 &    0.06 &    0.00 & $-$0.07 &    0.02 \\
  Na\,I & 6154.226 &    0.03 &    0.04 &    0.00 &    0.09 & $-$0.01 & $-$0.04 &    0.10 & $-$0.01 &    0.04 & $-$0.03 &    0.02 &    0.02 &    0.04 & $-$0.03 \\
  Na\,I & 6160.747 & $-$0.03 & $-$0.02 & $-$0.03 &    0.05 & $-$0.03 & $-$0.03 &    0.01 & $-$0.05 &    0.04 & $-$0.07 &    0.05 & $-$0.05 & $-$0.03 & $-$0.03 \\
  Mg\,I & 5183.000 & $-$0.03 & $-$0.04 & $-$0.05 &    0.05 & $-$0.03 & $-$0.05 & $-$0.06 & $-$0.08 &    0.04 & $-$0.10 &    0.00 & $-$0.02 &    0.04 &    0.01 \\
  Mg\,I & 5711.088 &    0.00 &    0.03 &    0.01 &    0.09 &    0.02 &    0.03 &    0.06 & $-$0.02 &    0.08 & $-$0.06 &    0.05 &    0.04 &    0.07 &    0.04 \\
  Al\,I & 6696.185 & $-$0.07 & $-$0.07 & $-$0.06 & $-$0.04 & $-$0.04 & $-$0.01 & $-$0.01 & $-$0.04 &    0.07 & $-$0.02 &    0.05 &    0.03 &    0.02 & $-$0.01 \\
  Al\,I & 6698.673 &    0.00 &    0.03 &    0.03 &    0.05 & $-$0.06 &    0.01 & $-$0.04 & $-$0.06 &    0.10 &    0.01 &    0.07 & $-$0.02 &    0.09 & $-$0.03 \\
  Si\,I & 5665.555 & $-$0.06 & $-$0.03 & $-$0.10 &    0.01 & $-$0.05 & $-$0.08 & $-$0.05 & $-$0.08 &    0.01 & $-$0.08 & $-$0.03 & $-$0.06 & $-$0.08 &    0.03 \\
  Si\,I & 5690.425 & $-$0.05 & $-$0.01 & $-$0.10 &    0.00 & $-$0.02 &    0.00 &    0.03 & $-$0.03 &    0.06 & $-$0.04 &    0.06 &    0.03 & $-$0.01 & $-$0.03 \\
  Si\,I & 6125.021 & $-$0.07 & $-$0.05 & $-$0.04 &    0.01 & $-$0.01 &    0.03 &    0.03 & $-$0.03 &    0.00 & $-$0.03 & $-$0.02 & $-$0.01 &    0.01 &    0.03 \\
  Si\,I & 6131.852 & $-$0.09 & $-$0.04 & $-$0.05 & $-$0.02 & $-$0.05 & $-$0.07 & $-$0.02 & $-$0.09 &    0.06 & $-$0.07 &    0.02 & $-$0.06 & $-$0.03 &    0.03 \\
  Si\,I & 6721.848 & $-$0.05 & $-$0.04 & $-$0.04 &    0.06 & $-$0.01 & $-$0.06 & $-$0.04 & $-$0.04 &    0.02 &    0.00 & $-$0.02 & $-$0.04 & $-$0.03 &    0.00 \\
   S\,I & 6046.027 &    0.04 &    0.00 & $-$0.06 &    0.10 & $-$0.07 & $-$0.13 & $-$0.06 & $-$0.01 & $-$0.05 &    0.08 & $-$0.11 & $-$0.08 &    0.02 &    0.02 \\
   S\,I & 6052.674 &    0.00 &    0.07 & $-$0.08 & $-$0.02 & $-$0.05 & $-$0.16 & $-$0.04 & $-$0.06 & $-$0.11 & $-$0.03 & $-$0.12 & $-$0.08 & $-$0.15 &    0.01 \\
   S\,I & 6757.171 &    0.05 &    0.09 &    0.06 & $-$0.02 &    0.09 & $-$0.03 &    0.08 &    0.05 &    0.00 &    0.09 &    0.05 &    0.00 & $-$0.01 &    0.00 \\
  Ca\,I & 5590.114 & $-$0.01 & $-$0.01 & $-$0.03 &    0.09 &    0.00 & $-$0.08 & $-$0.11 & $-$0.15 & $-$0.02 & $-$0.12 & $-$0.03 & $-$0.02 &    0.02 &    0.00 \\
  Ca\,I & 5867.562 & $-$0.02 &    0.03 & $-$0.02 &    0.08 & $-$0.02 & $-$0.05 &    0.04 & $-$0.04 &    0.07 & $-$0.02 & $-$0.03 & $-$0.08 &    0.06 & $-$0.04 \\
  Ca\,I & 6166.439 & $-$0.06 & $-$0.05 & $-$0.03 &    0.02 & $-$0.02 & $-$0.03 & $-$0.03 & $-$0.09 &    0.04 & $-$0.05 &    0.02 & $-$0.06 &    0.01 &    0.02 \\
  Ca\,I & 6169.042 &    0.04 &    0.06 &    0.01 &    0.12 &    0.04 &    0.01 & $-$0.01 & $-$0.06 &    0.06 &    0.00 &    0.04 & $-$0.02 &    0.05 &    0.01 \\
 Sc\,II & 5684.202 & $-$0.13 & $-$0.03 & $-$0.06 & $-$0.01 & $-$0.07 & $-$0.07 & $-$0.03 & $-$0.13 & $-$0.09 & $-$0.10 & $-$0.01 & $-$0.08 & $-$0.10 & $-$0.02 \\
 Sc\,II & 6245.637 & $-$0.05 & $-$0.01 & $-$0.09 &    0.00 & $-$0.06 & $-$0.07 & $-$0.05 & $-$0.08 &    0.07 & $-$0.08 &    0.01 & $-$0.10 & $-$0.06 &    0.01 \\
  Ti\,I & 5219.702 & $-$0.04 & $-$0.04 & $-$0.05 &    0.00 & $-$0.06 & $-$0.09 & $-$0.07 & $-$0.10 &    0.04 & $-$0.10 &    0.03 & $-$0.07 &    0.00 &    0.05 \\
  Ti\,I & 6126.216 & $-$0.09 & $-$0.10 & $-$0.05 &    0.05 & $-$0.09 &    0.05 &    0.02 & $-$0.06 &    0.04 & $-$0.08 &    0.04 & $-$0.01 &    0.04 &    0.04 \\
  Ti\,I & 6258.102 & $-$0.03 & $-$0.07 & $-$0.07 &    0.01 & $-$0.01 & $-$0.07 & $-$0.07 & $-$0.10 &    0.02 & $-$0.09 &    0.03 &    0.03 &    0.03 &    0.02 \\
 Ti\,II & 5490.690 & $-$0.09 & $-$0.07 & $-$0.08 & $-$0.06 & $-$0.06 & $-$0.02 & $-$0.06 & $-$0.10 & $-$0.01 & $-$0.10 & $-$0.08 & $-$0.07 & $-$0.07 &    0.03 \\
 Ti\,II & 6491.561 & $-$0.06 & $-$0.02 & $-$0.09 & $-$0.01 & $-$0.02 & $-$0.03 & $-$0.07 & $-$0.10 &    0.02 & $-$0.11 & $-$0.03 & $-$0.09 & $-$0.12 & $-$0.02 \\
   V\,I & 5670.853 &    0.02 &    0.01 & $-$0.11 &    0.02 &    0.05 & $-$0.01 &    0.05 & $-$0.10 &    0.06 & $-$0.11 &    0.05 & $-$0.02 &    0.07 &    0.02 \\
   V\,I & 6090.214 & $-$0.07 & $-$0.09 & $-$0.07 &    0.01 & $-$0.11 &    0.07 &    0.03 & $-$0.07 & $-$0.04 & $-$0.15 &    0.06 & $-$0.07 &    0.10 &    0.02 \\
  Cr\,I & 5238.961 & $-$0.04 & $-$0.02 &    0.05 & $-$0.06 & $-$0.03 & $-$0.01 & $-$0.09 & $-$0.03 &    0.12 & $-$0.09 &    0.06 & $-$0.03 &    0.06 & $-$0.06 \\
  Cr\,I & 5287.178 & $-$0.05 & $-$0.13 & $-$0.05 &    0.01 & $-$0.05 &    0.01 &    0.00 & $-$0.05 &    0.11 &    0.00 &    0.00 & $-$0.03 &    0.08 &    0.02 \\
  Mn\,I & 5377.637 & $-$0.08 & $-$0.05 & $-$0.11 &    0.03 & $-$0.05 & $-$0.02 &    0.04 & $-$0.09 &    0.02 & $-$0.08 &    0.13 & $-$0.02 &    0.01 &    0.01 \\
  Fe\,I & 5295.312 & $-$0.01 &    0.02 & $-$0.03 &    0.01 &    0.01 & $-$0.06 &    0.04 & $-$0.02 &    0.16 & $-$0.03 &    0.05 &    0.01 &    0.03 & $-$0.02 \\
  Fe\,I & 5522.446 & $-$0.07 & $-$0.03 & $-$0.10 &    0.00 & $-$0.04 & $-$0.11 &    0.00 & $-$0.11 &    0.00 & $-$0.17 & $-$0.05 & $-$0.04 &    0.00 &    0.04 \\
  Fe\,I & 6151.617 &    0.02 &    0.02 &    0.03 &    0.08 &    0.03 & $-$0.02 &    0.02 & $-$0.08 &    0.02 & $-$0.07 &    0.08 & $-$0.04 &    0.01 &    0.00 \\
  Fe\,I & 6229.226 &    0.05 &    0.02 & $-$0.03 &    0.07 &    0.03 &    0.08 &    0.04 &    0.03 &    0.12 &    0.02 &    0.12 &    0.02 &    0.10 &    0.01 \\
  Fe\,I & 6498.938 & $-$0.05 & $-$0.04 & $-$0.06 &    0.13 &    0.03 & $-$0.10 & $-$0.02 & $-$0.08 &    0.13 & $-$0.08 &    0.01 &    0.06 &    0.01 &    0.01 \\
 Fe\,II & 5414.073 & $-$0.04 & $-$0.04 & $-$0.12 & $-$0.01 & $-$0.13 & $-$0.17 & $-$0.18 & $-$0.16 & $-$0.10 & $-$0.16 & $-$0.09 & $-$0.16 & $-$0.19 &    0.01 \\
  Co\,I & 5342.695 & $-$0.06 &    0.02 & $-$0.02 &    0.06 &    0.03 & $-$0.02 &    0.04 & $-$0.03 &    0.04 & $-$0.11 &    0.04 &    0.02 &    0.07 & $-$0.03 \\
  Co\,I & 5530.774 & $-$0.06 & $-$0.01 & $-$0.08 &    0.08 & $-$0.04 &    0.02 &    0.05 & $-$0.01 &    0.03 & $-$0.08 & $-$0.05 & $-$0.05 & $-$0.01 &    0.04 \\
  Co\,I & 5647.234 & $-$0.03 & $-$0.02 & $-$0.03 &    0.14 & $-$0.11 & $-$0.01 & $-$0.06 & $-$0.08 & $-$0.02 & $-$0.15 &    0.03 & $-$0.06 & $-$0.07 &    0.06 \\
  Ni\,I & 5589.357 & $-$0.05 & $-$0.01 & $-$0.05 &    0.00 & $-$0.08 & $-$0.07 & $-$0.07 & $-$0.13 & $-$0.05 & $-$0.14 &    0.01 & $-$0.07 & $-$0.08 &    0.00 \\
  Ni\,I & 6086.276 & $-$0.03 &    0.02 & $-$0.05 &    0.05 &    0.01 & $-$0.07 &    0.00 & $-$0.05 &    0.07 & $-$0.05 &    0.11 &    0.05 & $-$0.02 & $-$0.03 \\
  Ni\,I & 6130.130 & $-$0.08 & $-$0.04 & $-$0.07 &    0.00 & $-$0.07 & $-$0.11 & $-$0.04 & $-$0.14 & $-$0.03 & $-$0.09 & $-$0.01 & $-$0.07 & $-$0.06 &    0.02 \\
  Ni\,I & 6204.600 & $-$0.04 & $-$0.04 & $-$0.05 &    0.00 & $-$0.04 & $-$0.04 &    0.04 & $-$0.03 &    0.06 & $-$0.05 &    0.01 & $-$0.07 & $-$0.01 &    0.02 \\
  Ni\,I & 6223.981 &    0.00 &    0.00 &    0.02 &    0.04 & $-$0.03 &    0.02 &    0.02 & $-$0.13 &    0.06 &    0.04 &    0.06 & $-$0.04 &    0.02 & $-$0.02 \\
  Ni\,I & 6772.313 & $-$0.05 &    0.01 &    0.00 &    0.02 & $-$0.04 & $-$0.05 & $-$0.03 & $-$0.09 &    0.06 & $-$0.08 &    0.00 & $-$0.06 & $-$0.04 &    0.00 \\
  Cu\,I & 5218.197 & $-$0.08 & $-$0.04 & $-$0.08 &    0.05 & $-$0.09 & $-$0.14 & $-$0.12 & $-$0.13 &    0.03 & $-$0.14 & $-$0.07 & $-$0.06 & $-$0.07 & $-$0.01 \\
  Cu\,I & 5220.066 &    0.03 &    0.00 &    0.00 &    0.03 & $-$0.08 &    0.03 & $-$0.10 &    0.02 &    0.14 & $-$0.09 &    0.06 &    0.10 &    0.08 &    0.03 \\
  Y\,II & 4900.120 & $-$0.09 & $-$0.06 & $-$0.08 & $-$0.03 & $-$0.02 & $-$0.03 & $-$0.06 & $-$0.12 &    0.00 & $-$0.10 &    0.01 & $-$0.03 & $-$0.03 & $-$0.02 \\
  Y\,II & 5087.416 & $-$0.08 & $-$0.05 & $-$0.10 & $-$0.01 & $-$0.01 & $-$0.02 & $-$0.07 & $-$0.09 & $-$0.02 & $-$0.14 &    0.08 & $-$0.11 & $-$0.15 & $-$0.01 \\
 \hline
  \end{tabular}
  \label{T4}
  \end{tiny}
 \end{table*} 

\begin{table*}[htbp]
  \caption{Abundance differences as a result of a \teff\ shift of +50\,K, a \logg\ shift of +0.05,
           a \vmic\ shift of +0.2\,kms$^{-1}$ and effects on individual abundances due to line scatter. 
           The fundamental parameters of the reference star M67-1194 were kept constant.}
  \begin{tabular}[h]{lrrr|rrr|rrr|rrr}
        &        & $\Delta$ \teff\ &         &        & $\Delta$ \logg\  &        &  & $\Delta$ \vmic\  & & & $\sigma$(lines) & \\
Elem    & MS1221 & TO1783 & SG1258 &  MS1221 & TO1783 & SG1258 & MS1221  & TO1783 & SG1258 & MS1221 & TO1783 & SG1258\\ 
  \hline \hline
  Li\,I &   0.04 &   0.04 &   0.04 &    0.00 &   0.00 &   0.00 & $-$0.01 &   0.00 &   0.00 &  $---$ &  $---$ & $---$ \\
   C\,I &$-$0.03 &$-$0.02 &$-$0.02 &    0.01 &   0.02 &   0.02 & $-$0.01 &   0.00 &$-$0.01 &  $---$ &  $---$ &  $---$\\
   O\,I &$-$0.02 &$-$0.02 &$-$0.01 &    0.01 &   0.02 &   0.02 &    0.00 &   0.00 &   0.00 &  0.16  &  0.16  &  0.06 \\
  Na\,I &   0.02 &   0.02 &   0.02 &    0.00 &   0.00 &$-$0.01 &    0.00 &$-$0.01 &$-$0.02 &  0.04  &  0.03  &  0.05 \\
  Mg\,I &   0.04 &   0.04 &   0.04 & $-$0.01 &$-$0.01 &$-$0.02 & $-$0.01 &$-$0.01 &$-$0.02 &  0.05  &  0.05  &  0.04 \\
  Al\,I &   0.01 &   0.02 &   0.02 &    0.00 &   0.00 &   0.00 &    0.00 &   0.00 &$-$0.01 &  0.07  &  0.01  &  0.04 \\
  Si\,I &   0.02 &   0.02 &   0.02 &    0.00 &   0.00 &   0.00 & $-$0.01 &$-$0.01 &$-$0.01 &  0.02  &  0.03  &  0.04 \\
   S\,I &$-$0.02 &$-$0.02 &$-$0.02 &    0.01 &   0.01 &   0.02 &    0.00 &$-$0.01 &   0.00 &  0.05  &  0.06  &  0.05 \\
  Ca\,I &   0.03 &   0.03 &   0.03 & $-$0.01 &   0.00 &   0.00 & $-$0.04 &$-$0.03 &$-$0.04 &  0.05  &  0.05  &  0.03 \\
 Sc\,II &   0.01 &   0.00 &   0.00 &    0.02 &   0.02 &   0.02 & $-$0.02 &$-$0.02 &$-$0.02 &  0.01  &  0.04  &  0.01 \\
     Ti &   0.03 &   0.03 &   0.03 &    0.01 &   0.01 &   0.01 & $-$0.01 &$-$0.01 &$-$0.02 &  0.03  &  0.02  &  0.05 \\
   V\,I &   0.04 &   0.04 &   0.05 &    0.00 &   0.00 &   0.00 & $-$0.01 &   0.00 &   0.00 &  0.07  &  0.02  &  0.04 \\
  Cr\,I &   0.03 &   0.03 &   0.02 &    0.00 &   0.00 &   0.00 &    0.00 &   0.00 &   0.00 &  0.08  &  0.01  &  0.00 \\
  Mn\,I &   0.03 &   0.03 &   0.03 &    0.00 &   0.00 &   0.00 & $-$0.01 &$-$0.01 &$-$0.02 &  $---$ &  $---$ &  $---$\\
     Fe &   0.03 &   0.03 &   0.03 &    0.00 &   0.00 &   0.00 & $-$0.02 &$-$0.01 &$-$0.02 &  0.03  &  0.07  &  0.08 \\
  Co\,I &   0.04 &   0.04 &   0.04 &    0.00 &   0.00 &   0.00 & $-$0.01 &$-$0.01 &   0.00 &  0.02  &  0.04  &  0.04 \\
  Ni\,I &   0.03 &   0.02 &   0.03 &    0.00 &   0.00 &   0.00 & $-$0.01 &$-$0.01 &$-$0.01 &  0.03  &  0.05  &  0.05 \\
  Cu\,I &   0.02 &   0.03 &   0.03 &    0.00 &   0.00 &   0.00 & $-$0.02 &$-$0.02 &$-$0.02 &  0.03  &  0.11  &  0.11 \\
  Y\,II &   0.02 &   0.01 &   0.01 &    0.02 &   0.02 &   0.02 & $-$0.06 &$-$0.07 &$-$0.08 &  0.01  &  0.02  &  0.06 \\
  \hline
  \end{tabular}
  \label{T5}
 \end{table*} 

\end{landscape}
\end{onecolumn}

\end{appendix}

\end{document}